\documentclass[]{spie}
\usepackage[]{fontenc}
\usepackage[latin1]{inputenc}
\usepackage{amsmath}
\usepackage{graphicx}

\makeatletter

\providecommand{\LyX}{L\kern-.1667em\lower.25em\hbox{Y}\kern-.125emX\@}

\usepackage{latexsym}
\newtheorem{dfn}{Definition}
\def\qed{\hfill $\Box$ \vskip 3pt}

\makeatother
\begin{document}

\title{Quantum Modeling\date{}}

\author{Darin Goldstein, Computer Science Department, Cal State Long
Beach, \textit{daring@cecs.csulb.edu}}

\maketitle
\begin{abstract}
We present a modification of Simon's Algorithm \cite{Sim94,Sim97}
that in some cases is able to fit experimentally obtained data to
appropriately chosen trial functions with high probability. Modulo
constants pertaining to the reliability and probability of success
of the algorithm, the algorithm runs using only $O(polylog(|Y|))$
queries to the quantum database and $O(polylog(|X|,|Y|))$ elementary
quantum gates where $|X|$ is the size of the experimental data set
and $|Y|$ is the size of the parameter space. We discuss heuristics
for good performance, analyze the performance of the algorithm in
the case of linear regression, both one-dimensional and multidimensional,
and outline the algorithm's limitations.
\end{abstract}

\section{Introduction}\label{sec:Introduction}

Assume that two experiments are run to determine the parameters governing
a physical phenomenon. If very few data points are obtained, then
it is difficult to determine the best parameters via a cursory examination
of a data plot.

\includegraphics[  scale=0.4]{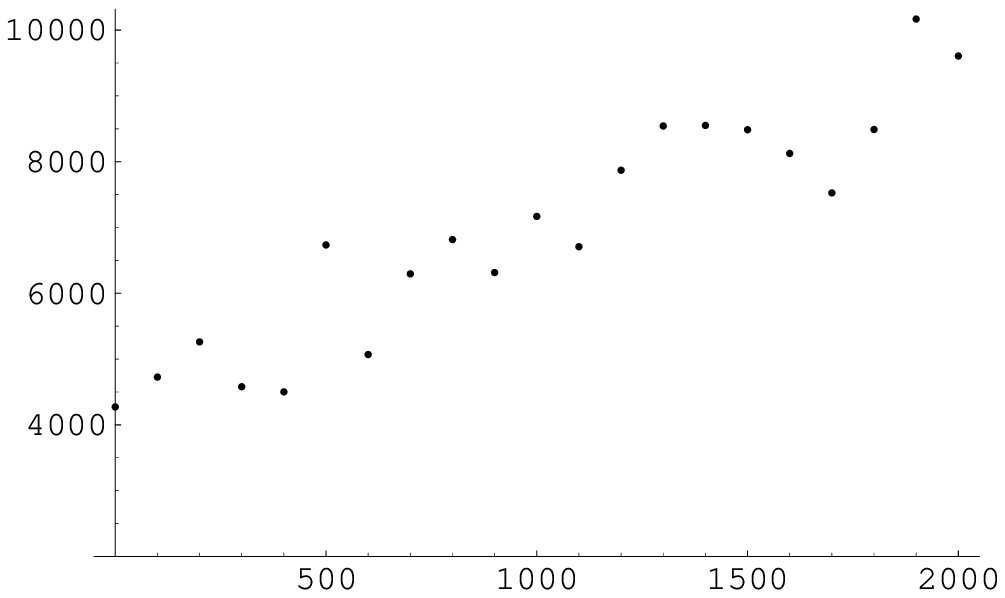}

In this case, one might abandon the plot and use a
computer to quickly calculate the best parameter range using standard
techniques. On the other hand, if numerous experimental data points
are obtained and a plot of the data is saturated with such points,
one would have far less trouble detecting a good fit to the data;
of course, given such a large quantity of data, the same computer
would most likely take far longer for its analysis (assuming that
no sampling techniques are used to pare down the data).

\includegraphics[  scale=0.4]{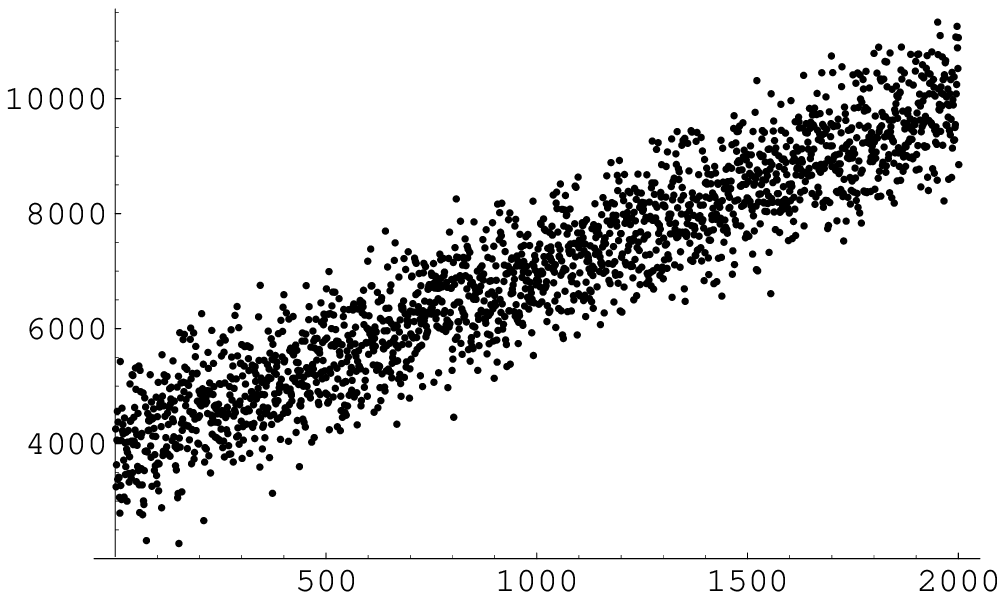}

There appears
to be an relationship between the ease with
which computers and people analyze data.

The point of this mental exercise is to note that classical computers
do not have the human ability to {}``eyeball'' a graph, to capture
the shape of a graph by evaluating the data it contains {}``all at
once''. One essential difference between classical (and even classical
probabilistic) computation and quantum computation is the ability
of the quantum computer to evaluate a function at more than one point
at a time. For example, given a function $f$ on a single bit, a quantum
computer has the ability to evaluate $f$ at the point $|x\rangle =\frac{1}{\sqrt{2}}|0\rangle +\frac{1}{\sqrt{2}}|1\rangle $,
an $x$ value that has no meaning classically%
\footnote{Throughout this paper, we will use the notation of the vector formulation
of quantum computation as outlined in Chuang and Nielsen \cite{ChuangNeilsen}.%
}, in order to realize both $f(0)$ and $f(1)$ in the resulting state.
The simplest and earliest use of this interesting property in tandem
with quantum interference is Deutsch's algorithm \cite{Deu85} which
is able to extract global properties of $f$ in fewer evaluations
than was possible classically. In this paper, we will show how classical
artificial intelligence tasks, in particular, fitting a mathematical
model to given experimental data, may in some cases be sped up via
the use of quantum computation and quantum data representation.

Throughout the paper, we use $f$ consistently to represent the experimental
data. We perform a physical experiment and obtain classical data that
we then store in a quantum database indexed by the independent variables
of the experiment. Thus, the action of $f$ on a basis vector $|x\rangle $
is to return the value of the experimental data when the independent
variable was $x$. More generally, if a superposition of basis vectors
is entered into the quantum database then the database returns the
appropriate superposition of outputs. In other words, we define the
quantum database by the following action%
\footnote{For completeness, we present an alternate physical scenario in which
our algorithm would apply equally well. We are given a black box that
takes as input a vector of input and scratch qubits and implements
the function $f$. The black box performs some unknown quantum operations
on the qubits and outputs the results of the computation of $f$ and
possibly other output scratch qubits that remain in safe storage.
We assume that there exists a basis (and, without loss of generality,
one should assume it is the standard basis) such that the action of
the experiment function $f$ is as follows. For any basis vector $|x\rangle $,
$f:|x\rangle \otimes |b\rangle \mapsto |x\rangle \otimes |b\oplus f(x)\rangle $
where $|f(x)\rangle $ is also a basis vector. Our goal is to model
the black box experiment $f$ by some classical function.%
}:\[
f:(\sum _{x\in X}c_{x}|x\rangle )\otimes |b\rangle \mapsto \sum _{x\in X}c_{x}|x\rangle \otimes |b\oplus f(x)\rangle \]
(This database formulation is roughly the same computational model
as the one used in Grover's quantum database search \cite{Grover96,Grover97}.)
We will use the variable $y$ to represent the parameter vector and
$g_{y}$ to represent the parameterized function to which we wish
to fit the data. We will assume that $g_{y}$ is a classical function
implemented with quantum gates. (For example, if $f(x)=3\sqrt{x}$,
then a good guess for $g_{y}$ is $g_{y}(x)=y\sqrt{x}$. Note that
$x$ and $y$ need not always be single numbers. See Section \ref{section:algorithm}.) 

In Section \ref{sec:Mathematical-Preliminaries}, we review some mathematical
preliminaries and give a rigorous grounding to the notion of what
it means for two functions to have a {}``similar shape''. In this
section, we also discuss previous research and give several references
on the topic of multidimensional functional optimization. The precise
mathematical implementation of the various functions, the main algorithm,
and a discussion of the query/time-complexity are presented in Section
\ref{section:algorithm}. A rigorous mathematical analysis of the
performance of the algorithm in the case of one-dimensional and multivariable
linear regression is given in Section \ref{sec:LinearAnalysis}. An
example of the algorithm running on a nonlinear sample input as well
as certain tedious mathematical justifications are in the Appendix,
Section \ref{section:appendix}.

\section{Mathematical Preliminaries}\label{sec:Mathematical-Preliminaries}

\subsection{The Shape Measure}\label{sub:The-Shape-Measure}

Assume that we are given a function $f$ and another function $g_{y}$
that is supposed to approximate $f$. Both are defined on the domain
$X$. We ask the following question: For a given value of $y$, how
good an approximation to $f$ is $g_{y}$? Usually, when such a question
is asked, the {}``accuracy'' of the approximation is quantified
in a single number. The most common measure is the sum of the squared
distances:\[
L_{2}(f,g_{y})=\sum _{x\in X}|f(x)-g_{y}(x)|^{2}\]
 Analysis of this $L_{2}$ measure under the assumption that $g_{y}(x)=yx$
leads to the methods of linear regression \cite{BurdenFaires89},
probably the most standard data fitting technique in use today.

Unfortunately, analysis of the $L_{2}$ measure is not always easy.
In the case where the function $f$ is given by experimentally obtained
data, the expression given by $L_{2}(f,g_{y})$ has $|X|$ terms in
it. If $|X|$ is large, $y$ is a vector with many elements, or the
parameterized function $g_{y}$ is not well-behaved (as opposed to
linear regression), minimizing $L_{2}(f,g_{y})$ can be quite a difficult
and resource-consuming problem. Global function optimization has been
an active field of research for years. In particular, we refer the
reader to the brief review of continuous global optimization by Pint$\acute{\textrm{e}}$r
\cite{Pint96}. Numerous Internet sites offer publicly available global
optimization software that take as input a user-defined black-box
function. These programs attempt to minimize the number of function
evaluations necessary to obtain the global maximum/minimum. We cite
as examples those listed in the software review by Mongeau et al.
\cite{mongeau98comparison}: adaptive simulated annealing \cite{Ing93},
clustering algorithms \cite{Boender82,Csendes}, genetic algorithms
\cite{Jelasity}, multi-level random search \cite{Kunt95}, Monte-Carlo
based random search \cite{Zheng95}, and various combinations \cite{Luksan96}.

In order to get a relatively small value for $L_{2}$, it is necessary
to choose $y$ such that $g_{y}$ is vertically close to the function
$f$. When data is collected experimentally, most of the time scientists
are interested in the shape of the graph rather than the vertical
intercept because the shape has more physical significance. For this
reason, we define a measure to compete with the $L_{2}$ standard,
the shape measure%
\footnote{The function $Q$ that we define in this section is not a metric in
the strict mathematical sense. It is, however, intuitively a good
measure of separation of functions relative to a given sensitivity
$M$. We will show in the Appendix, Section \ref{sub:The-Pseudometric-QStar},
that with high probability, there exists a value of $M$ for which
there is a simple modification of $Q$ that leads to a true metric
with high probability and we bound it between reasonable modifications
of $L_{1}$ and $L_{2}$. In the rest of the paper, we continue to
use $Q$ as our measure of interest even though it is not technically
a metric.%
} $Q$, as follows.\begin{dfn}\[
Q(f,g_{y},M)=|\sum _{x\in X}\frac{e^{\frac{2\pi i}{M}(f(x)-g_{y}(x))}}{|X|}|^{2}\]
\end{dfn}The parameter $M$ should be thought of as the {}``sensitivity''
of the measure $Q$: the lower the value of $M$, the more sensitive
the measure is to small perturbations in the shapes of the graphs
of $f$ and $g_{y}$. The measure $Q$ is somewhat counterintuitive
in the sense that the more similar the shapes of $f$ and $g$ are,
the \emph{larger} the value of $Q(f,g_{y},M)$ and vice versa.

We can think of this function intuitively as follows. Imagine a compass
on a unit circle initially pointing East. For each value of $x$,
move the compass pointer around the unit circle clockwise if $f(x)>g_{y}(x)$
and counterclockwise if $f(x)<g_{y}(x)$. The greater the difference
between $f(x)$ and $g_{y}(x)$, the further around the compass you
move. If we can arrange it so that the most we can move around the
compass is, for example, under $\frac{\pi }{2}$ radians, then if
$f(x)$ is much greater than $g_{y}(x)$, the compass will be pointing
close to North. Similarly, if $f(x)$ is much less than $g_{y}(x)$,
the compass will be pointing South. Given the compass direction for
each value of $x$, walk one unit for every direction that the compass
points. Squaring your final distance from the origin is what $Q(f,g_{y},M)$
measures. 

Note that $Q(f,g_{y},M)$ gives the same distance no matter which
direction we arbitrarily call {}``East''. This special property
corresponds to the invariance under vertical translation that we need.

\begin{lemma}\label{lemma:translationalinvariance}For any real numbers
$k_{1}$ and $k_{2}$ and any sensitivity $M$, $Q(f,g_{y},M)=Q(f+k_{1},g_{y}+k_{2},M)$.\end{lemma}

We now justify the use of the term {}``sensitivity'' for the parameter
$M$. Intuitively, the following lemma states that as the sensitivity
$M$ increases, the two functions are differentiated with certainty.
As the sensitivity decreases, they tend to look approximately the
same.

\begin{dfn}For any real number $a$, let \[
frac(a)=\left\{ \begin{array}{cc}
 a-\lfloor a\rfloor  & if\, a\ge 0\\
 a-\lceil a\rceil  & if\, a<0\end{array}
\right.\]
 (Intuitively, $frac(a)$ is just the fractional part of the number
$a$.)\end{dfn}

\begin{lemma}\label{lemma:Mbounds}For any functions $f\neq g_{y}$,
assume that as $M\rightarrow 0^{+}$, $frac(\frac{f-g_{y}}{M})$ becomes
uniformly distributed in $(-\frac{1}{2},\frac{1}{2})$. Then\[
\lim _{M\rightarrow \infty }Q(f,g_{y},M)=1\]
and\[
\lim _{M\rightarrow 0^{+}}Q(f,g_{y},M)\approx \frac{1}{|X|}\]
\end{lemma}

\emph{Proof}. The first limit is clear from elementary calculus even
without the assumption. We sketch the estimation for the second limit.

Firstly, note that as $M\rightarrow 0^{+}$, by the assumption $frac(\frac{f-g_{y}}{M})$
is uniformly distributed in $(-\frac{1}{2},\frac{1}{2})$. Recalling
the formula for $Q(f,g_{y},M)$, it is clear that for small enough
$M$, each step $e^{\frac{2\pi i}{M}(f(x)-g_{y}(x))}$ is uniformly
distributed around the unit circle. By elementary probability theory,
the expected value of the squared distance for a random walk in two
dimensions is $|X|$. Thus, taking into account the $|X|$ in the
denominator, we get the estimate $\frac{1}{|X|}$ for $Q(f,g_{y},M)$
for small enough $M$.\qed

\subsection{\label{subsection:parameterspectrum}The Parameter Spectrum}

When attempting to search for parameters that fit within a given mathematical
model for a scientific phenomenon, rather than confine the final value
of the parameterization to the single {}``best'' value (as is usually
done with linear regression and other methods of model fitting), we
propose a different approach. Consider the spectrum of all possible
parameter values. Obviously, some parameter values will better fit
the experimental data than others. If the mathematical model is good
enough, there should be a small range of parameter values that are,
in some sense, much better than all the others. Identifying this small
range, the range of parameter values that are {}``good enough'',
given a predetermined error tolerance, is the goal of our parameter
search.

Assume we are given the usual approximation function $g_{y}(x)$ where
$x$ is the vector of independent variables and $y$ is the parameter
vector. Let $Y$ be the set of all possible values for the vector
$y$. For a given parameter range $Y'\subseteq Y$, we define the
\emph{parameter ratio} $r(Y',Y)$ as follows. \begin{dfn}\[
r(Y',Y)=\frac{\sum _{y\in Y'}Q(f,g_{y},M)}{\sum _{y\in Y}Q(f,g_{y},M)}\]
\end{dfn}Note that $\forall Y'\subseteq Y,0\le r(Y',Y)\le 1$. Recall
that the higher the value of $Q$, the better the shape fit between
$f$ and $g_{y}$. Thus if the shape fit for the parameter values
in $Y'$ are better than the other values in $Y$, this ratio $r(Y',Y)$
should be significantly higher than other possible ranges.

In the algorithm to follow in Section \ref{subsection:trimming},
we will reduce the parameter space based on the values of
the parameter ratios. We assume that there is a predetermined threshold
associated with these parameter ratios in the following sense. If
the parameter ratio of the parameter range $Y'\subseteq Y$ is greater
than, for example, 65\%, we might say that we are certain enough that
the optimal parameter range is somewhere in $Y'$. If we can determine
a $Y'$ with a parameter ratio 65\%, all further searches of the parameter
space can be restricted to the space $Y'$. Obviously, the number
65\% is arbitrary and left up to the individual scientist running
the experiment.

\section{The Algorithm}\label{section:algorithm}

We will continue using the functions $f$ and $g_{y}$ as defined
previously. We assume that $f$ and $g_{y}$ both evaluate to nonnegative
integers for every value of the vectors $x$ and $y$. (This restriction
may be removed via simple transformations of the functions $f$ and
$g_{y}$, but we will assume it here for simplicity.) We assume that
we can evaluate the functions $f$ and $g_{y}$ out to any given fixed
precision.

Consider the following simple two-dimensional example: We have a quantum
oracle that evaluates the function $f(x_{1},x_{2})=5x_{1}+16x_{2}+noise$
where the $noise$ term refers to a small random variable contributed
by the environment or other factors. We have deduced that $f$ evaluates
a function that looks roughly like a plane. Thus, our approximation
function will be $g_{(y_{1},y_{2})}(x_{1},x_{2})=y_{1}x_{1}+y_{2}x_{2}$.
Our aim, of course, is to find the best range for the parameters $y_{1}$
and $y_{2}$. Let us assume that we know that the parameters can only
take values less than 32. We will therefore dedicate 5 bits for each
and start our search space out with $(y_{1},y_{2})\in [0,31]\times [0,31]$.
Note that we could have, given more analysis of the data, restricted
the variable $y_{1}$ to only 3 bits without losing any pertinent
information. However, if we suspected that the parameter $y_{1}$
included a fractional part (i.e. if we suspected that $f(x_{1},x_{2})=5.25x_{1}+16x_{2}+noise$),
we could keep the extra bits and use them for additional precision
by modifying the approximation function as follows: $g_{(y_{1},y_{2})}(x_{1},x_{2})=\frac{y_{1}}{4}x_{1}+y_{2}x_{2}$.
Negative numbers and other mathematical possibilities can be handled
in a similar way.

In order to make the discussion of Section \ref{sec:Introduction}
more precise, we assume that we have a black box which computes both
the functions $f$ and $g_{y}$ to $N$ digits of binary precision
in the following sense. On basis vectors, our evaluator for $f$ computes
(where $\oplus $ is binary addition mod $2^{N}$)\[
|x_{1},x_{2}\rangle \otimes |b\rangle \mapsto |x_{1},x_{2}\rangle \otimes |b\oplus f(x_{1},x_{2})\rangle \]
where each $|\cdot \rangle $ represents a bit vector of the appropriate
size. ($|b\rangle $ will have $N$ binary digits, as specified.)
We then extend this to the entire vector space by linearity. This
is a well-known model of function evaluation and is the basis of most
popular quantum algorithms including Shor's factoring algorithm \cite{Shor94,Shor97}
and quantum database search \cite{Grover96,Grover97}. On basis vectors,
our evaluator for $g$ computes (where $\ominus $ is binary subtraction
mod $2^{N}$)\[
|x_{1},x_{2},y_{1},y_{2}\rangle \otimes |b\rangle \mapsto |x_{1},x_{2},y_{1},y_{2}\rangle \otimes |b\ominus g_{(y_{1},y_{2})}(x_{1},x_{2})\rangle \]
and we again extend this by linearity.

\subsection{\label{subsection:restrictions}Heuristics for Good Performance}

Though the algorithm presented here is provably fast and accurate
in some cases (see Section \ref{sec:LinearAnalysis}), we are not
optimistic enough to claim that this is so in every case. It is very
difficult to find rigorous general restrictions that will \emph{guarantee}
satisfactory performance. However, this section will present some
common-sense heuristics in the form of assumptions. Sections \ref{subsection:choosingN}
through \ref{sub:The-Overall-Algorithm} present the algorithm itself.

\begin{enumerate}
\item \label{item:firstassumption}We should assume that, for all practical
purposes, $\exists y'\in Y$ such that\[
f(x)=g_{y'}(x)+noise\]
We can think of $noise$ as a uniform random variable taking values
in $[-\epsilon ,\epsilon ]$ for some small $\epsilon >0$ or as a
standard normal random variable with relatively small variance. (If
the mathematical model is off by a quasi-insignificant term, it may
be safely incorporated into the $noise$ term.)
\item \label{item:secondassumption}We should assume that the $noise$ term
is small relative to the difference in model parameter. $\forall y\neq y'\in Y$
(where $y'$ is defined as in assumption \ref{item:firstassumption}),\[
Var(noise)\ll Var(f-g_{y})\]
If the noise from the environment drowns out the mathematical model,
it will be very difficult to come up with good parameters because
\emph{any} choice of parameters will likely be a bad one%
\footnote{Classically, before one can determine how bad the noise is, the parameters
need to be chosen first. Without knowledge of what the model is, one
can't determine how much of the data is noise. It is a quantum feature
of the algorithm that if the environmental noise drowns out the experimental
data \emph{even before we know what is and isn't noise}, the algorithm
will not be able to make a good choice for $Y'$. So, if there happens
to be no {}``good'' parameter choice, the algorithm simply won't
make a choice.%
}.
\item \label{item:thirdassumption}We should assume that the mathematical
model $g_{y}$ we ascribe to the quantum oracle $f$ is, in a sense,
{}``continuous'' in the parameter $y$; that a small change in the
$y$ parameter should not produce too large a change in the quantum
distance measure $Q$. On the other hand, we also need to assume that
$Q$ differentiates between good parameter choices and bad ones. Thus,
we should assume that there exist constants $a>0$ and $b<1$ such
that $\forall g_{y}\exists M$ such that $a\le E(Q(g_{y},g_{\cdot },M))\le b$.
\end{enumerate}
We will use these heuristics as assumptions below in Section \ref{sec:LinearAnalysis}
when we analyze the performance of the algorithm on linear regression.

\subsection{\label{subsection:choosingN}Choosing the Sensitivity}

This entire subsection should be thought of as a function within a
greater procedure. It determines, if possible, the best value of the
sensitivity for $Q$ for use in further calculations.

Below, we use the notation $X$ to refer to the set of all possible
values for the independent variable $|x\rangle $. We initialize the
value of $N$ to be $\lceil \log _{2}\max _{x\in X,y\in Y}\{f(x),g_{y}(x)\}\rceil $,
the maximum number of digits that the quantum evaluator needs to evaluate
either $f$ or $g_{y}$. Throughout this section, we let the function
$FFT$ be the quantum fast Fourier Transform \cite{ChuangNeilsen}
whose action on basis vectors is $FFT:|b\rangle \mapsto \sum _{k=0}^{2^{N}-1}\frac{e^{\frac{2\pi ibk}{2^{N}}}}{\sqrt{2^{N}}}|k\rangle $.

\begin{enumerate}
\item We first create the superposition%
\footnote{This step basically follows the procedure in Shor \cite{Shor97}.%
}\[
\sum _{x\in X}\frac{e^{\frac{-2\pi if(x)}{2^{N}}}}{\sqrt{|X|}}|x\rangle \]
in the following way. Start with $|x\rangle =|0\ldots 0\rangle $
and $|b\rangle =|0\ldots 01\rangle $. (Assume that $|b\rangle $
contains $N$ total binary digits.)\[
|0\ldots 0\rangle \otimes |0\ldots 01\rangle \mapsto \]
\[
H^{\otimes \log |X|}\otimes FFT:\sum _{x\in X}\frac{|x\rangle }{\sqrt{|X|}}\otimes \sum _{k=0}^{2^{N}-1}\frac{e^{\frac{2\pi ik}{2^{N}}}}{\sqrt{2^{N}}}|k\rangle \mapsto \]
\[
f:\sum _{x\in X}\frac{e^{\frac{-2\pi if(x)}{2^{N}}}}{\sqrt{|X|}}|x\rangle \otimes \sum _{k=0}^{2^{N}-1}\frac{e^{\frac{2\pi ik}{2^{N}}}}{\sqrt{2^{N}}}|k\rangle \]
Because this system is separable, we can discard the second expression,
and this leaves us with the system we want.
\item We now introduce another system representing the parameter states.
Let $Y$ be the set that represents the current search space for the
parameter vector $|y\rangle $. Our final goal is a system in the
form\[
\sum _{y\in Y}\sum _{x\in X}\frac{e^{\frac{2\pi i(g_{y}(x)-f(x))}{2^{N}}}}{\sqrt{|X||Y|}}|x\rangle \otimes |y\rangle \]
We initialize $|y\rangle =|0\ldots 0\rangle $. Again we will make
use of $|b\rangle =|0\ldots 01\rangle $ with $N$ binary digits.
Using the system we created in the previous step, we get\[
\sum _{x\in X}\frac{e^{\frac{-2\pi if(x)}{2^{N}}}}{\sqrt{2^{N}}}|x\rangle \otimes |0\ldots 0\rangle \otimes |0\ldots 01\rangle \mapsto \]
\[
I^{\otimes \log |X|}\otimes H^{\otimes \log |Y|}\otimes FFT:\]
\[
\sum _{x\in X}\frac{e^{\frac{-2\pi if(x)}{2^{N}}}}{\sqrt{|X|}}|x\rangle \otimes \sum _{y\in Y}\frac{|y\rangle }{\sqrt{|Y|}}\otimes \sum _{k=0}^{2^{N}-1}\frac{e^{\frac{2\pi ik}{2^{N}}}}{\sqrt{2^{N}}}|k\rangle \mapsto \]
\[
g:\]
\[
\sum _{x\in X}\frac{e^{\frac{2\pi i(g_{y}(x)-f(x))}{2^{N}}}}{\sqrt{|X|}}|x\rangle \otimes \sum _{y\in Y}\frac{|y\rangle }{\sqrt{|Y|}}\otimes \sum _{k=0}^{2^{N}-1}\frac{e^{2\pi ik}}{\sqrt{2^{N}}}|k\rangle \]
 Again separability allows us to ignore the $|b\rangle $ qubits and
leaves us with the desired system.
\item Finally, we perform a Hadamard transformation on the $|x\rangle $
qubit system. So we get\[
\sum _{z\in X}\sum _{y\in Y}\sum _{x\in X}\frac{e^{\frac{2\pi i(g_{y}(x)-f(x))}{2^{N}}}}{|X|\sqrt{|Y|}}(-1)^{x\cdot z}|z\rangle \otimes |y\rangle \]

\item \label{item:zprob}The final step in this procedure is to measure
the $|z\rangle $ qubits. Based on the state expansion outlined above,
the probability that $|z\rangle =|0\ldots 0\rangle $ is\[
\frac{1}{|Y|}\sum _{y\in Y}|\sum _{x\in X}\frac{e^{\frac{2\pi i(g_{y}(x)-f(x))}{2^{N}}}}{|X|}|^{2}=\]
\[
\frac{1}{|Y|}\sum _{y\in Y}Q(f,g_{y},2^{N})=E(Q(f,g_{\cdot },2^{N}))\]
The exact value of this probability can be estimated to arbitrary
accuracy with high probability by repeated trials using the above
steps. (Make sure to keep those experiments such that $|z\rangle =|0\ldots 0\rangle $
as they will likely be useful later. Also note the similarity of the
final expression to Assumption \ref{item:thirdassumption} of Section
\ref{subsection:restrictions}.) We can now make our decision as to
whether we have chosen an appropriate value of the sensitivity value
$2^{N}$. If our estimate of the $|z\rangle =|0\ldots 0\rangle $
measurement probability is not within $[\frac{1}{10},\frac{3}{5}]$
(arbitrarily chosen constants%
\footnote{These constants were chosen because of their good performance in numerical
experiments by the author, but their exact values do not really matter.
If the range is too great, it will simply take longer for other parts
of the process to produce good results. We only desire that the probability
that we measure $|z\rangle =|0\ldots 0\rangle $ be bounded above
and below by constants. See Assumption \ref{item:thirdassumption}
of Section \ref{subsection:restrictions}.%
}), we proceed through the following list of checks. (Note that we
may fall into category (c) after we have checked (a) and (b). The
final check is not mutually exclusive from the other two.)

\begin{enumerate}
\item If the $|z\rangle =|0\ldots 0\rangle $ measurement probability is
below $\frac{1}{10}$ (set above), we double the value of $2^{N}$
(increasing $N$ by 1) and try again. If this happens, it means that
the sensitivity was so low that we are only capturing the behavior
of the $noise$ in the function $f$ and not so much the differences
in the function $g_{y}$ as we vary the parameter $y$.
\item If the $|z\rangle =|0\ldots 0\rangle $ measurement probability is
above $\frac{3}{5}$ (set above), we halve the value of $2^{N}$ (decreasing
$N$ by 1 if possible%
\footnote{If we reach this point and it is not possible to decrease $N$ any
further (i.e. $N=1$), we terminate the function and report that no
sensitivity value will work.%
}) and try again. In this case, the sensitivity was artificially high.
The value of $2^{N}$ was dwarfing the both the parameter differences
of $g_{y}$ \emph{and} the $noise$ term in $f$.
\item If the $|z\rangle =|0\ldots 0\rangle $ measurement probability exceeds
the boundaries in the opposite direction of the previous experiment,
we terminate the algorithm and declare that the parameter space $Y$
cannot be improved upon. In other words, the predictive value of the
mathematical model is not significantly improved by making the parameter
space smaller than it already is.
\end{enumerate}
\end{enumerate}
Once we get the value of $N$ that we desire from our above experiments
by assuring ourselves within a predetermined tolerance that $P(|z\rangle =|0\ldots 0\rangle )\in [\frac{1}{10},\frac{3}{5}]$,
we can continue on to the next phase.

\emph{Remark}. As soon as the value of $\frac{2^{N}}{2}$ (and almost
certainly even $\frac{2^{N}}{4}$) approaches $|noise|$, the algorithm
is likely to terminate: If $\frac{noise}{2^{N}}$ is a uniform random
variable taking values between $-\frac{1}{2}$ and $\frac{1}{2}$,
then $E(Q(f,g_{\cdot },2^{N}))=\frac{1}{|X|}$. See Lemma \ref{lemma:Mbounds}
for justification.

\subsection{\label{subsection:trimming}Trimming the Parameter Space}

In this section, we describe how the parameter space may be trimmed
to half its size with high probability. Again, as with the above section,
we assume that this section is a function within a larger procedure.
Assume below that we have chosen a field $y_{i}$ to work with within
the parameter vector $y$. The ultimate goal of this function is to
cut the possible values of $y_{i}$ by half correctly with high probability.

\begin{enumerate}
\item Determine the optimal sensitivity $2^{N}$ using the function described
in Section \ref{subsection:choosingN}.
\item Collect several systems (the exact number to be determined by the
error tolerance) that have already been measured with $|z\rangle =|0\ldots 0\rangle $.
Recalling that the state of the system before the $|z\rangle =|0\ldots 0\rangle $
measurement is\[
\sum _{z\in X}\sum _{y\in Y}\sum _{x\in X}\frac{e^{\frac{2\pi i(g_{y}(x)-f(x))}{2^{N}}}}{|X|\sqrt{|Y|}}(-1)^{x\cdot z}|z\rangle \otimes |y\rangle \]
the final state of the system after the $|z\rangle =|0\ldots 0\rangle $
measurement is\[
\sum _{y\in Y}\sum _{x\in X}\frac{\frac{e^{\frac{2\pi i(g_{y}(x)-f(x))}{2^{N}}}}{|X|\sqrt{|Y|}}}{\sqrt{P(|z\rangle =|0\ldots 0\rangle )}}|y\rangle \]
where $P(|z\rangle =|0\ldots 0\rangle )$ is defined as in Step \ref{item:zprob}
of Section \ref{subsection:choosingN}.
\item \label{item:trimstep}If possible, split the parameter space of $y_{i}$
into 4 equal pieces by considering the 2 most significant bits of
$y_{i}$. (The case where $y_{i}$ only has a single bit is analogous
and will be explicitly outlined below.) For $j\in [0,3]$, let $Y_{ij}$
be the subset of the parameter space such that the two most significant
digits of $y_{i}$ equals $j$. For each of the systems collected
above, measure the parameter $y_{i}$ and keep track of which $Y_{ij}$
the measurement falls into. The probability of measuring $y_{i}$
in the a given $Y_{ij}$ is equal to\[
\frac{\sum _{y\in Y_{ij}}|\sum _{x\in X}\frac{e^{\frac{2\pi i(g_{y}(x)-f(x))}{2^{M}}}}{|X|\sqrt{|Y|}}|^{2}}{P(|z\rangle =|0\ldots 0\rangle )}\]
\[
=\frac{\sum _{y\in Y_{ij}}Q(f,g_{y},M)}{\sum _{y\in Y}Q(f,g_{y},M)}=r(Y_{ij},Y)\]
We will consider narrowing our search for the optimal parameter value
to either $Y_{i0}\cup Y_{i1}$, $Y_{i1}\cup Y_{i2}$, or $Y_{i2}\cup Y_{i3}$%
\footnote{At this point, the astute reader might be asking himself why we bother
to split the parameter space in quarters when it looks like we may
be able to split it in half and speed up the algorithm by a factor
of 2. Let us assume that $y_{i}$ can be between 0 and 31 and the
optimal value is 15 but both 15 and 16 are acceptable parameter values
for our model (see Section \ref{subsection:parameterspectrum}). We
split the parameter space into 4 parts: $[0,7],[8,15],[16,23],[24,31]$.
Clearly, $[8,15]$ should have a large probability mass, but so might
$[16,23]$ if 16 is an acceptable parameter value as well.%
}. Recalling the discussion of Section \ref{subsection:parameterspectrum},
we take into account just how different we need the parameter value
to be in order to perform this narrowing of the parameter space%
\footnote{One point should be clarified before proceeding further: If the parameter
ratio of one of the three half-spaces does not meet the threshold
for further narrowing of the parameter space, there are two possibilities
for why: (a) The parameter spectrum need not be narrowed any further
because the data fits the model approximately the same way for each
half-space. In this case, no narrowing of the search is desirable.
(b) The mathematical model/search space have not been chosen well.
For example, the search space might be so large that good parameter
values in a half space are drowned out by the large number of bad
choices.%
}. Once we have made our choice as to which part of the space we are
interested in examining, we modify the function $g_{y}$, replacing
$y_{i}$ with the appropriate linear transformation (See Section \ref{sub:The-Overall-Algorithm}.)
and reducing the number of search bits in $y_{i}$ by one. We then
continue on with the algorithm.\\
\\
It remains to discuss what occurs when the search space cannot be
split into 4 equal pieces. In this case, $y_{i}$ is either 0 or 1.
We simply run the analogous experiment described above for the quarter
case and choose the correct half of the space based on the given threshold.
Once we have made the final decision if possible, $y_{i}$ is eliminated
as a parameter.\\
\\
If it turns out that we cannot meet the threshold for trimming the
parameter space, we mark $y_{i}$ finished as a parameter and continue
on to $y_{i+1}$.
\end{enumerate}

\subsection{The Overall Algorithm\label{sub:The-Overall-Algorithm}}

The algorithm amounts to repeated application of the procedure outlined
in Section \ref{subsection:trimming}.

While the parameter space continues to shrink in size: for each parameter
in succession, perform the trimming described above in Section \ref{subsection:trimming}.
Whenever a parameter is trimmed, the function $g_{y}$ is updated:
if we are working on the variable $y_{i}$ and choose the half space
$Y_{i0}\cup Y_{i1}$ (resp. $Y_{i2}\cup Y_{i3}$), then the function
$g_{y}$ automatically inserts a 0 (resp. 1) for the high order bit
of $y_{i}$ and the number of input bits for $y_{i}$ gets reduced
by 1. If we choose the half space $Y_{i1}\cup Y_{i2}$ and $y_{i}$
contains $b$ bits, then we let the high order bit of $y_{i}$ be
0, reduce the number of input bits in $y_{i}$ to $b-1$ (similar
to the $Y_{i0}\cup Y_{i1}/Y_{i2}\cup Y_{i3}$ cases), and then before
evaluation of $g_{y}$ perform the $b$ bit addition $y_{i}\mapsto y_{i}+2^{b-2}-1$.
(The single bit case is analogous.) Once the parameter space ceases
to shrink in size, output the reduced parameter spaces for each of
the parameters.

The Appendix (Section \ref{sub:A-Nonlinear-Example}) contains a fully
worked out nonlinear example of this algorithm.

\subsection{The Query/Time Complexity}

Modulo various constants depending on the error tolerance and threshold
values, the algorithm presented above requires $O(polylog(|X|,|Y|))$
elementary quantum operations (as defined in \cite{ChuangNeilsen})
and $O(polylog(|Y|))$ evaluations of the functions $f$ and $g_{y}$.

By way of comparison, we examine the classical running time for finding
the best fit for a trial function using $|X|$ experimental data points
and parameter space of size $|Y|$. The standard technique is to minimize
the function $L_{2}(f,g_{y})$. Note that simply writing down the
function $L_{2}(f,g_{y})$ requires time $\Omega (|X|)$ (if we do
no probabilistic sampling of the search space). Minimizing $L_{2}(f,g_{y})$
with respect to the parameter vector $y$ requires either solving
the equations $\nabla L_{2}(f,g_{y})=0$ or minimizing $L_{2}(f,g_{y})$
directly using a global optimization protocol. (Writing down the equations
$\nabla L_{2}(f,g_{y})=0$ requires time proportional to both $|X|$
and the number of parameters in the parameter vector $y$. Note that
in the case where the number of parameters in the parameter vector
$y$ is large, say $\Omega (|Y|^{c})$ for some small $c<1$, it takes
time $\Omega (|X||Y|^{c})$ just to write down the equations to be
solved.) If the functions $g_{y}$ are sufficiently nasty, this could
necessitate solving equations numerically using a probabilistic method,
genetic algorithm, or other previously mentioned sampling technique.
If we have to sample $\Theta (|Y|^{k})$ times for some constant $k<1$,
then the classical running time for finding the best fit parameter
vector $y$ could reach $\Theta (|X||Y|^{k})$. This is exponentially
worse than the quantum algorithm presented here. If probabilistic
sampling from the search space is done, the quantum algorithm and
classical probabilistic algorithms still perform at asymptotically
similar speeds.

In conclusion, the inherently quantum nature of interference allows
potentially exponential speedup for sufficiently well-chosen parameterized
functions. Constructive interference takes place when the data matches
the function, and the quantum coefficients interfere destructively
when the parameters provide a poor fit.

\subsection{Limitations and Areas of Future Research}

The algorithm we present here has some obvious limitations. In this
section, we present a simple example to illustrate its shortcomings.

Consider the function that is 0 everywhere except at a single unknown
point $a$ where it takes the value 1. We wish to discover the value
of $a$ by fitting this function to a function of the form\[
g_{y}(x)=\left\{ \begin{array}{cc}
 0 & if\, x\neq y\\
 1 & if\, x=y\end{array}
\right.\]
Clearly, we want the end value of the parameter $y$ to equal $a$.

At this point, note that this situation violates one of our heuristics
for good performance, specifically Assumption \ref{item:thirdassumption}
in Section \ref{subsection:restrictions}. These functions are not
effectively separated by the quantum distance measure $Q$; on the
other hand, $L_{2}$ is 1 for every value $y\neq a$ and 0 if $y=a$.
We do not expect our quantum algorithm to perform well. Indeed, it
cannot possibly perform well: this situation is equivalent to the
database search problem. There is a well-known result that any quantum
algorithm that solves the database search problem with high probability
must make $\Omega (\sqrt{N})$ queries to the database where $N$
is the number of elements in the database \cite{bennett94strengths}.
In our case, the number of elements is equal to $|Y|$; because our
algorithm makes only $O(polylog(|Y|))$ {}``queries'', it cannot
possibly find the correct value in time with high probability.

One interesting possible area of research we leave open is to determine
the family of functions for which the algorithm is able to perform
correctly with high probability. Section \ref{sec:LinearAnalysis}
below will guarantee good behavior from the algorithm in the linear
and multilinear case. However, it appears that determining where the
separation between {}``well-behaved'' families of functions (such
as the linear and multilinear cases) and {}``databases'' is a very
difficult problem. In Ambainis \cite{ambainis00quantum}, it is claimed that
{}``the unordered search problem provides an abstract model for NP-complete
problems.'' Indeed, it appears that one might just as well ask the
question of where the separation occurs between boolean functions
for which there exists a polynomial-time satisfiability algorithm
and those that are truly intractable.

\section{Analysis of Linear Regression}\label{sec:LinearAnalysis}

This section will present a rigorous analysis of necessary and sufficient
conditions for good behavior in the case where we use the algorithm
to perform linear regression.

\subsection{One-Dimensional Linear Regression}

\begin{dfn}Using the same notation as in the previous sections, we
will assume that $Y=\{-2^{K-1},\ldots ,2^{K-1}-1\}$ and $X=\{0,\ldots ,2^{L}-1\}$.
Then $\log _{2}|X|=L$ and $\log _{2}|Y|=K$. $N$ has the same sensitivity
meaning as it did above in Section \ref{section:algorithm}. Finally,
let $N=L+K+r$ where $r$ is a constant whose optimal value will be
worked out in the analysis to follow.\end{dfn}

We will assume $f(x)=y'x+noise$ for some $y'\in Y$ (see Section
\ref{subsection:restrictions}, Assumption \ref{item:firstassumption}).
So $g_{y}(x)=yx$. We will assume that $noise$ is a Gaussian random
variable with small variance. (The mean is irrelevant because of the
translational invariance of $Q$.) Recall that\[
P(|z\rangle =|0\ldots 0\rangle )=\frac{1}{|X|^{2}|Y|}\sum _{y\in Y}|\sum _{x\in X}e^{2\pi i\frac{g_{y}(x)-f(x)}{2^{N}}}|^{2}\]
\[
=\frac{1}{2^{2L+K}}\sum _{y=-2^{K-1}}^{2^{K-1}-1}|\sum _{x=0}^{2^{L}-1}e^{\frac{2\pi i(y-y')x}{2^{N}}}e^{\frac{2\pi i(noise)}{2^{N}}}|^{2}\]
We look to find a lower bound for this expression. Recalling Assumption
\ref{item:secondassumption} of Section \ref{subsection:restrictions},
our first approximation is that $Var(\frac{noise}{2^{N}})\ll 1\Rightarrow $(by
translational invariance)$e^{\frac{2\pi i(noise)}{2^{N}}}\approx 1$.
That reduces our expression to\[
\approx \frac{1}{2^{2L+K}}\sum _{y=-2^{K-1}}^{2^{K-1}-1}|\sum _{x=0}^{2^{L}-1}e^{\frac{2\pi i(y-y')x}{2^{N}}}|^{2}\]
\[
=\frac{1}{2^{2L+K}}\sum _{y=-2^{K-1}}^{2^{K-1}-1}|\frac{1-e^{\frac{2\pi i(y-y')2^{L}}{2^{N}}}}{1-e^{\frac{2\pi i(y-y')}{2^{N}}}}|^{2}\]
Let $u=\frac{y-y'}{2^{K}}$.\[
=\frac{1}{2^{2L+K}}\sum _{u=-\frac{1}{2}-\frac{y'}{2^{K}}}^{\frac{1}{2}-\frac{1}{2^{K}}-\frac{y'}{2^{K}}}(\frac{\sin (2^{L+K-N}\pi u)}{\sin (2^{K-N}\pi u)})^{2}\]
Letting $y^{*}=\frac{y'}{2^{K}}\in [-\frac{1}{2},\frac{1}{2}]$, recalling
that $N=L+K+r$, and approximating the sum with an integral, we get\[
\approx \int _{-\frac{1}{2}-y^{*}}^{\frac{1}{2}-y^{*}}(\frac{\sin (2^{-r}\pi u)}{2^{L}\sin (2^{-L-r}\pi u)})^{2}du\]
As $L$ gets large, $2^{L}$ gets very large. We make the approximation
that $2^{L}\rightarrow \infty $. So taking this limit and letting
$w=2^{-r}\pi u$, we get\[
\approx \frac{2^{r}}{\pi }\int _{\frac{\pi }{2^{r}}(-\frac{1}{2}-y^{*})}^{\frac{\pi }{2^{r}}(\frac{1}{2}-y^{*})}(\frac{\sin w}{w})^{2}dw\]
Note that this expression is independent of all variables except $r$
and $y^{*}$. The following is a plot of $P(|z\rangle =|0\ldots 0\rangle )$.
The $r$ value is on the $x$-axis (front) and the $y^{*}$ value
is on the $y$-axis (right).

\includegraphics[  scale=0.4]{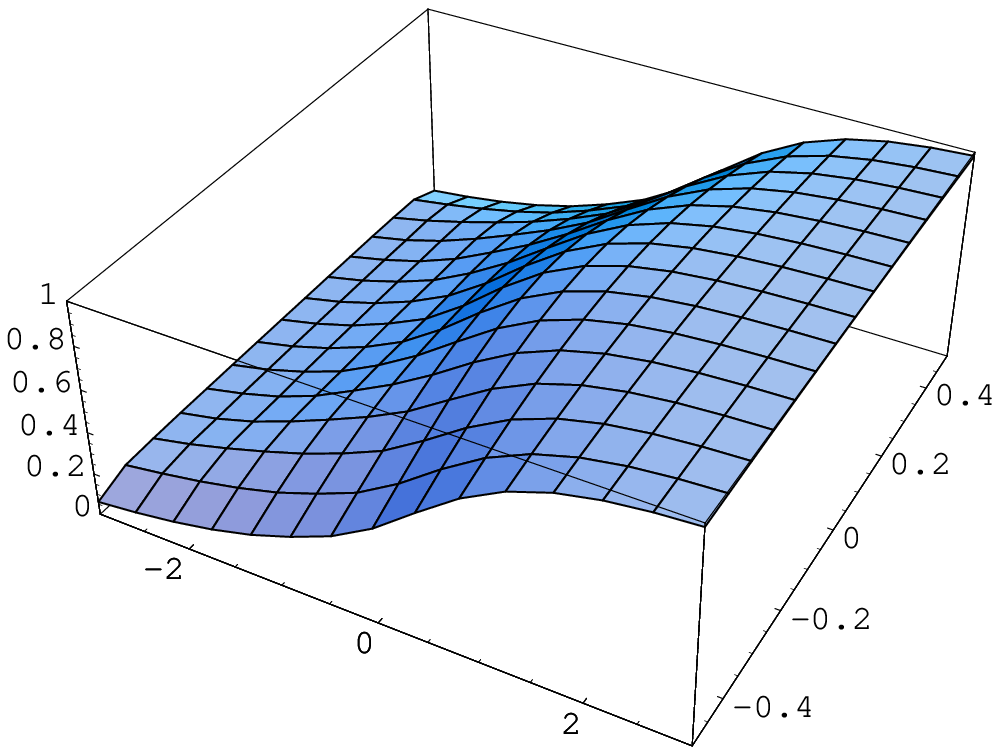}

Given this expression for $P(|z\rangle =|0\ldots 0\rangle )$, it
is now easy to calculate the probabilities of the three relevant half
spaces:\[
\frac{\frac{2^{r}}{\pi }\int _{\frac{\pi }{2^{r}}(-\frac{1}{2}-y^{*})}^{\frac{\pi }{2^{r}}(0-y^{*})}(\frac{\sin w}{w})^{2}dw}{\frac{2^{r}}{\pi }\int _{\frac{\pi }{2^{r}}(-\frac{1}{2}-y^{*})}^{\frac{\pi }{2^{r}}(\frac{1}{2}-y^{*})}(\frac{\sin w}{w})^{2}dw}\]
\[
\frac{\frac{2^{r}}{\pi }\int _{\frac{\pi }{2^{r}}(-\frac{1}{4}-y^{*})}^{\frac{\pi }{2^{r}}(\frac{1}{4}-y^{*})}(\frac{\sin w}{w})^{2}dw}{\frac{2^{r}}{\pi }\int _{\frac{\pi }{2^{r}}(-\frac{1}{2}-y^{*})}^{\frac{\pi }{2^{r}}(\frac{1}{2}-y^{*})}(\frac{\sin w}{w})^{2}dw}\]
\[
\frac{\frac{2^{r}}{\pi }\int _{\frac{\pi }{2^{r}}(0-y^{*})}^{\frac{\pi }{2^{r}}(\frac{1}{2}-y^{*})}(\frac{\sin w}{w})^{2}dw}{\frac{2^{r}}{\pi }\int _{\frac{\pi }{2^{r}}(-\frac{1}{2}-y^{*})}^{\frac{\pi }{2^{r}}(\frac{1}{2}-y^{*})}(\frac{\sin w}{w})^{2}dw}\]
We can calculate the maximum of these three half spaces for any given
values of $r$ and $y^{*}$. Below is a plot of this maximum.

\includegraphics[  scale=0.4]{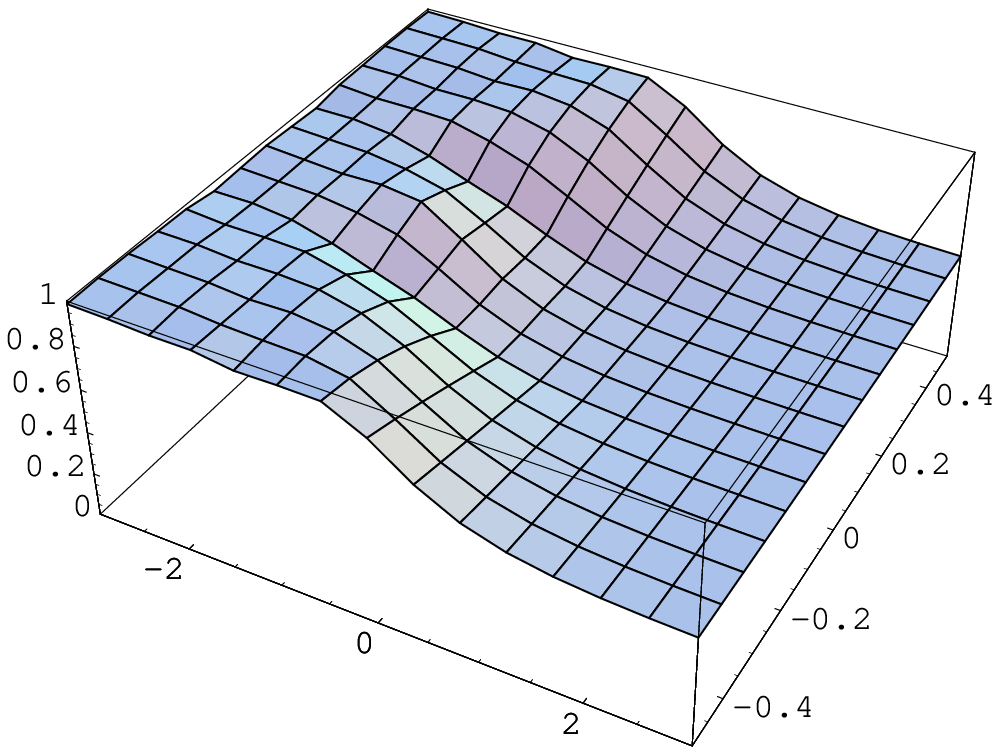}

By examining these plots, we notice that if we choose $r=-1$, then
$P(|z\rangle =|0\ldots 0\rangle )>20\%$ and the half space with the
greatest probability must have weight at least 70\%. So rather than
performing the algorithm outlined in Section \ref{subsection:choosingN},
in the case of linear regression with a single variable, we simply
choose $N=L+K-1$.

\emph{Remark}. It is somewhat strange that the maximum value that
$f(x)-g_{y}(x)$ takes in this case requires $L+K$ bits to compute.
Our result above shows that we can \emph{ignore} the highest order
bit of the computation.

\subsection{Multivariate Linear Regression\label{sub:Multivariate-Linear-Regression}}

The analysis of the multivariable case parallels that of the one-dimensional.

\begin{dfn}Using the same notation as the previous sections, we assume
that $Y=\{-2^{K-1},\ldots ,2^{K-1}-1\}^{\times d}$ and $X=\{0,\ldots ,2^{L}-1\}^{\times d}$.
Let $N=L+K+r$ as before.\end{dfn}

We assume that $f(x)=\sum _{j=1}^{d}y_{j}'x_{j}+noise$ for some vector
$y'\in Y$ and $g_{y}(x)=\sum _{j=1}^{d}y_{j}x_{j}$. The random variable
$noise$ is still Gaussian with small variance. The results are as
follows.\[
P(|z\rangle =|0\ldots 0\rangle )\approx \prod _{j=1}^{d}(\frac{2^{r}}{\pi }\int _{\frac{\pi }{2^{r}}(-\frac{1}{2}-y_{j}^{*})}^{\frac{\pi }{2^{r}}(\frac{1}{2}-y_{j}^{*})}(\frac{\sin w}{w})^{2}dw)\]
where $\forall j\in \{1,\ldots ,d\},\, y_{j}^{*}=\frac{y_{j}'}{2^{K}}$.
The plot for the greatest of the three half spaces is \emph{exactly}
the same as the one-dimensional case. Thus, if we choose $r=-1$,
the probability that we measure $P(|z\rangle =|0\ldots 0\rangle )$
decreases exponentially with $d$; however, there is no decrease in
the relative weight of the half spaces. The mathematical justification
for these results can be found in the Appendix (Section \ref{sub:The-Multivariate-Mathematics}).

\section*{Acknowledgments}

The author would like to thank Tim Robertson, William
Murray, and Todd Ebert for helpful suggestions with both form
and content.

\bibliographystyle{plain}
\bibliography{quantumai}

\section{Appendix}\label{section:appendix}

\subsection{The Pseudometric $Q^{*}$\label{sub:The-Pseudometric-QStar}}

This section introduces a pseudo-metric $Q^{*}=1-\sqrt{Q}$ (for small
enough $M$). The function $1-\sqrt{Q}$ is a metric {}``with high
probability'' in the sense that the properties of a metric are satisfied
with high probability given appropriate conditions that we outline.
We introduce this function in order to relate the new measure $Q$
to other, more commonly used metrics like $L_{1}$ and $L_{2}$.

We must be careful about describing the metric space over which we
claim $Q^{*}$ as a pseudo-metric. Consider the space of all real-valued
functions $f(x)$. We identify two functions $f$ and $g$ as equivalent
if $f=g+k$ for some constant $k$. We claim that $Q^{*}$ is a metric
for the resulting space of equivalence classes. We denote by $\overline{f}$
the equivalence class of the function $f$. Note that by translational
invariance (Lemma \ref{lemma:translationalinvariance}), $Q^{*}$
is well-defined over this space.

\begin{lemma}The function $Q^{*}=1-\sqrt{Q}$ is a metric over the
set of translational equivalence classes of functions with high probability
for small enough values of $M$ and large enough $|X|$.\end{lemma}

\emph{Proof}. We need to check three conditions. Only the triangle
inequality is nontrivial. We need to show that for functions $f,g,$
and $h$, $(1-\sqrt{Q(f,g,M)})+(1-\sqrt{Q(g,h,M)})\ge 1-\sqrt{Q(f,h,M)}$.
Note that by Lemma \ref{lemma:Mbounds}, if $M$ gets small, all of
the $Q$ values approach $\frac{1}{|X|}$ with high probability. If
$|X|$ is large enough, the inequality is true by inspection.\qed

It is also possible to bound $Q^{*}$ between relevant modifications
of two well-known metrics, $L_{1}$ and $L_{2}$.

\begin{lemma}\label{lemma:bounds}Let\[
L_{1}^{*}(f,g,M)=\sum _{x\in X}\frac{|f(x)-g(x)|}{|X|M}\]
 and\[
L_{2}^{*}(f,g,M)=\sum _{x\in X}\frac{|f(x)-g(x)|^{2}}{|X|M^{2}}\]
 (Note that $L_{1}^{*}$ and $L_{2}^{*}$ are not well-defined on
the space of vertically equivalent functions. $L_{1}^{*}$ is an average
normalized distance and $L_{2}^{*}$ is an average normalized squared
distance.) Then there exists $M>0$ such that\[
2\pi L_{1}^{*}(f,g,M)+(1-\frac{\pi }{2})\le Q^{*}(\overline{f},\overline{g},M)\le 2\pi ^{2}L_{2}^{*}(f,g,M)\]
\end{lemma}

\emph{Proof}. Let $M_{1}=\max _{x\in X}4|f(x)-g(x)|$. Then we have
that $\forall x\in X,|\frac{f(x)-g(x)}{M_{1}}|\le \frac{1}{4}$. Given
this value for $M_{1}$, we now perform the random walk given by $Q$
described in Section \ref{sub:The-Shape-Measure}. An appropriate
vertical modification to the function $f$, say replacing $f$ with
$f+k_{1}$ for some constant $k_{1}$, will suffice to place the endpoint
of the random walk on the $x$-axis. If it is still true that $\forall x\in X,|\frac{f(x)-g(x)}{M_{1}}|\le \frac{1}{4}$,
then we let $M=M_{1}$. Otherwise, given the new function $f+k_{1}$,
we repeat the process which leads to $M_{2}$ and $f+k_{1}+k_{2}$
for some constant $k_{2}$. We continue evaluating $M_{i}$ until
we have found our value of $M$. Note that these steps must eventually
terminate because the size of $\max _{x\in X}|\frac{f(x)-g(x)}{M_{i}}|$
is decreasing substantially with every iteration.

Note that after the process of determining $M$ has completed, $f$
has been modified to $f+k_{1}+k_{2}+\ldots +k_{n}$ (if there are
$n$ iterations) and the random walk is guaranteed to terminate on
the $x$-axis. Thus, using standard approximations for $\sin (x)$
and $\cos (x)$, we get\[
Q(f,g,M)=|\sum _{x\in X}\frac{\cos (2\pi \frac{f(x)-g(x)}{M})}{|X|}|^{2}\]
\[
\ge |\sum _{x\in X}\frac{1-2\pi ^{2}(\frac{f(x)-g(x)}{M})^{2}}{|X|}|^{2}\]
\[
=(1-2\pi ^{2}L_{2}^{*}(f,g,M))^{2}\]
and\[
Q(f,g,M)=|\sum _{x\in X}\frac{\cos (2\pi \frac{f(x)-g(x)}{M})}{|X|}|^{2}\]
\[
=|\sum _{x\in X}\frac{\sin (\frac{\pi }{2}-2\pi |\frac{f(x)-g(x)}{M}|)}{|X|}|^{2}\]
\[
\le \frac{\pi ^{2}}{4}|1-4L_{1}^{*}(f,g,M)|^{2}\]
Algebra yields the resulting bounds.\qed

\emph{Remark}. Note that the bounds given in Lemma \ref{lemma:bounds}
are essentially tight. On the right, if $f=g$, then $L_{2}^{*}(f,g,M)=0\Rightarrow Q(f,g,M)=1$.
On the left, if $f$ is as far from $g$ as possible, then $L_{1}^{*}(f,g,M)=\frac{1}{4}\Rightarrow Q(f,g,M)=0$.

\subsection{The Multivariate Mathematics\label{sub:The-Multivariate-Mathematics}}

We will be using the same notation as that given in Section \ref{sub:Multivariate-Linear-Regression}.
Recall that\[
P(|z\rangle =|0\ldots 0\rangle )=\frac{1}{|X|^{2}|Y|}\sum _{y\in Y}|\sum _{x\in X}e^{2\pi i\frac{g_{y}(x)-f(x)}{2^{N}}}|^{2}\]
and if we assume that the effect of the $noise$ term is negligible,
we get\[
\approx \frac{1}{2^{2Ld+Kd}}\sum _{y_{1}=-2^{K-1}}^{2^{K-1}-1}\ldots \sum _{y_{d}=-2^{K-1}}^{2^{K-1}-1}|\sum _{x_{1}=0}^{2^{L}-1}\ldots \sum _{x_{d}=0}^{2^{L}-1}\prod _{j=1}^{d}e^{2\pi i\frac{(y_{j}-y_{j}')x_{j}}{2^{N}}}|^{2}\]
\[
=\frac{1}{2^{2Ld+Kd}}\sum _{y_{1}=-2^{K-1}}^{2^{K-1}-1}\ldots \sum _{y_{d}=-2^{K-1}}^{2^{K-1}-1}\prod _{j=1}^{d}|\sum _{x_{j}=0}^{2^{L}-1}e^{2\pi i\frac{(y_{j}-y_{j}')x_{j}}{2^{N}}}|^{2}\]
\[
=\prod _{j=1}^{d}\frac{1}{2^{2L+K}}\sum _{y_{j}=-2^{K-1}}^{2^{K-1}-1}|\sum _{x_{j}=0}^{2^{L}-1}e^{2\pi i\frac{(y_{j}-y_{j}')x_{j}}{2^{N}}}|^{2}\]
Now, we perform the same one-dimensional analysis for each $j$ term
to get the final result.\[
\approx \prod _{j=1}^{d}(\frac{2^{r}}{\pi }\int _{\frac{\pi }{2^{r}}(-\frac{1}{2}-y_{j}^{*})}^{\frac{\pi }{2^{r}}(\frac{1}{2}-y_{j}^{*})}(\frac{\sin w}{w})^{2}dw)\]
To see that the half space calculations turn out the same, note that
we only change the top two bits of a single variable at a time. Only
the $j$ term for that variable changes in the half space calculation.
All other variables cancel out.

\subsection{A Nonlinear Example\label{sub:A-Nonlinear-Example}}

This section performs the algorithm of Section \ref{section:algorithm}
on a specific example. We will use $f(x_{1},x_{2})=x_{1}^{2}+16x_{2}+constant+noise$
where $noise$ is a random variable that take values uniformly between
-30 and 30 and $constant$ is a vertical constant that we care nothing
about. Data has been collected from the domain $[0,31]\times [0,31]$.
The parameter function we use is $g(x_{1},x_{2},y_{1},y_{2})=y_{1}x_{1}^{2}+y_{2}x_{2}$.
We will use 3 bits for $y_{1}$ and 5 bits for $y_{2}$. We use a
threshold of 60\% when performing Step \ref{item:trimstep} of Section
\ref{subsection:trimming}. Our acceptable range for measuring $|z\rangle =|0\ldots 0\rangle $
will be $[\frac{1}{10},\frac{3}{5}]=[.1,.6]$. The following is a
3D graph of the function $f(x_{1},x_{2})$.

\vspace{0.3cm}
\begin{center}\includegraphics[  scale=0.4]{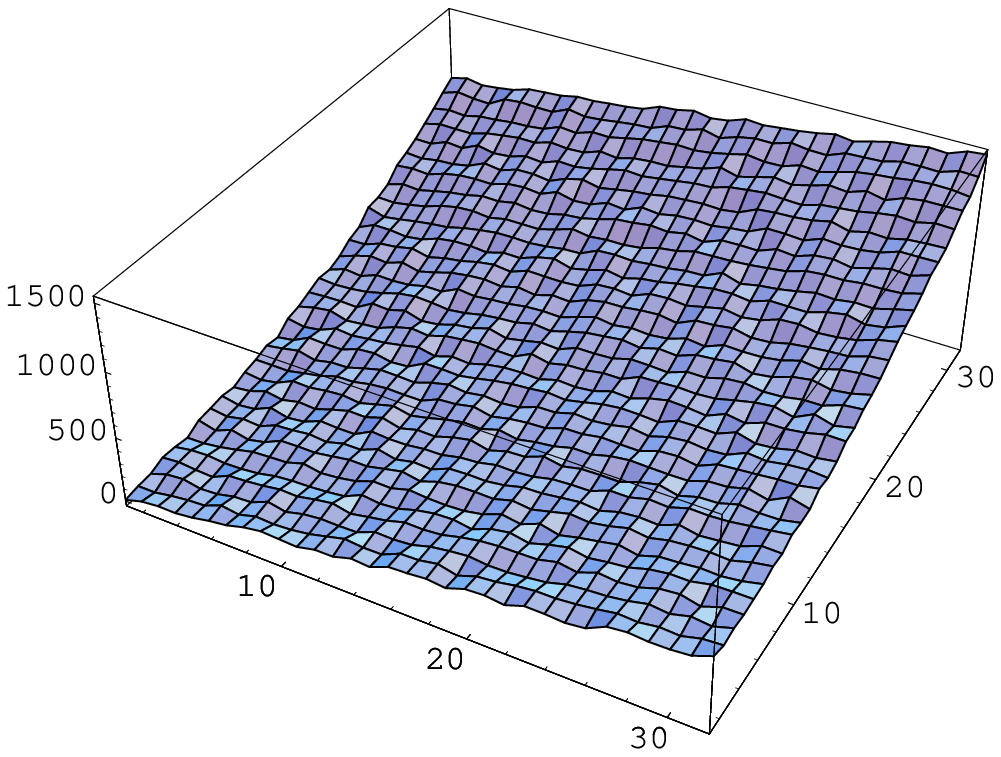}\end{center}
\vspace{0.3cm}

\begin{enumerate}
\item Begin by working on parameter $y_{1}\in [0,7]$.

\begin{enumerate}
\item We initialize $2^{N}=2^{11}\Rightarrow P(|z\rangle =|0\ldots 0\rangle )=0.265781\in [.1,.6]$.
This sensitivity is accepted.
\item The probabilities of measuring $y_{1}$ in the relevant ranges are
as follows:

\begin{itemize}
\item $P(y_{1}\in [0,1])=0.643622$
\item $P(y_{1}\in [2,3])=0.230168$
\item $P(y_{1}\in [4,5])=0.0778675$
\item $P(y_{1}\in [6,7])=0.0483422$\\
We accept the range $y_{1}\in [0,3]$ with total probability $0.87379$.
(tolerance 60\%)
\end{itemize}
\item Recalculating with $2^{N}=2^{11}\Rightarrow P(|z\rangle =|0\ldots 0\rangle )=0.448745\in [.1,.6]$.
This sensitivity is accepted.
\item The probabilities of measuring $y_{1}$ in the relevant ranges are
as follows:

\begin{itemize}
\item $P(y_{1}\in [0,0])=0.217404$
\item $P(y_{1}\in [1,1])=0.519182$
\item $P(y_{1}\in [2,2])=0.217218$
\item $P(y_{1}\in [3,3])=0.046196$\\
We accept the range $y_{1}\in [0,1]$ with total probability $0.736586$.
(tolerance 60\%)
\end{itemize}
\item Recalculating with $2^{N}=2^{11}\Rightarrow P(|z\rangle =|0\ldots 0\rangle )=0.448745\in [.1,.6]$.
This sensitivity is accepted.
\item We have narrowed $y_{1}$ to two possible values.

\begin{itemize}
\item $P(y_{1}=0)=0.295151$
\item $P(y_{1}=1)=0.704849$\\
We accept $y_{1}=1$. (tolerance 60\%)
\end{itemize}
\end{enumerate}
\item Working on parameter $y_{2}\in [0,31]$.

\begin{enumerate}
\item Recalculating with $2^{N}=2^{11}\Rightarrow P(|z\rangle =|0\ldots 0\rangle )=0.931922\not \in [.1,.6]$.
This sensitivity is not accepted.
\item Recalculating with $2^{N}=2^{10}\Rightarrow P(|z\rangle =|0\ldots 0\rangle )=0.764582\not \in [.1,.6]$.
This sensitivity is not accepted.
\item Recalculating with $2^{N}=2^{9}\Rightarrow P(|z\rangle =|0\ldots 0\rangle )=0.431486\in [.1,.6]$.
This sensitivity is accepted.
\item The probabilities of measuring $y_{2}$ in the relevant ranges are
as follows:

\begin{itemize}
\item $P(y_{2}\in [0,7])=0.0546121$
\item $P(y_{2}\in [8,15])=0.40125$
\item $P(y_{2}\in [16,23])=0.453112$
\item $P(y_{2}\in [24,31])=0.0910255$\\
We accept the range $y_{2}\in [8,23]$ with total probability $0.854362$.
(tolerance 60\%)
\end{itemize}
\item Recalculating with $2^{N}=2^{9}\Rightarrow P(|z\rangle =|0\ldots 0\rangle )=0.737291\not \in [.1,.6]$.
This sensitivity is not accepted.
\item Recalculating with $2^{N}=2^{8}\Rightarrow P(|z\rangle =|0\ldots 0\rangle )=0.375476\in [.1,.6]$.
This sensitivity is accepted.
\item The probabilities of measuring $y_{2}$ in the relevant ranges are
as follows:

\begin{itemize}
\item $P(y_{2}\in [8,11])=0.0398828$
\item $P(y_{2}\in [12,15])=0.371314$
\item $P(y_{2}\in [16,19])=0.475369$
\item $P(y_{2}\in [19,23])=0.113435$\\
We accept the range $y_{2}\in [12,19]$ with total probability $0.846683$.
(tolerance 60\%)
\end{itemize}
\item Recalculating with $2^{N}=2^{8}\Rightarrow P(|z\rangle =|0\ldots 0\rangle )=0.635818\not \in [.1,.6]$.
This sensitivity is not accepted.
\item Recalculating with $2^{N}=2^{7}\Rightarrow P(|z\rangle =|0\ldots 0\rangle )=0.206041\in [.1,.6]$.
This sensitivity is accepted.
\item The probabilities of measuring $y_{2}$ in the relevant ranges are
as follows:

\begin{itemize}
\item $P(y_{2}\in [12,13])=0.0170483$
\item $P(y_{2}\in [14,15])=0.300177$
\item $P(y_{2}\in [16,17])=0.511306$
\item $P(y_{2}\in [18,19])=0.171468$\\
We accept the range $y_{2}\in [14,17]$ with total probability $0.811483$.
(tolerance 60\%)
\end{itemize}
\item Recalculating with $2^{N}=2^{7}\Rightarrow P(|z\rangle =|0\ldots 0\rangle )=0.334398\in [.1,.6]$.
This sensitivity is accepted.
\item The probabilities of measuring $y_{2}$ in the relevant ranges are
as follows:

\begin{itemize}
\item $P(y_{2}\in [14,14])=0.115462$
\item $P(y_{2}\in [15,15])=0.254449$
\item $P(y_{2}\in [16,16])=0.336552$
\item $P(y_{2}\in [17,17])=0.293536$\\
We accept the range $y_{2}\in [16,17]$ with total probability $0.630089$.
(tolerance 60\%)
\end{itemize}
\item Recalculating with $2^{N}=2^{7}\Rightarrow P(|z\rangle =|0\ldots 0\rangle )=0.334398\in [.1,.6]$.
This sensitivity is accepted.
\item We have narrowed $y_{2}$ to two possible values.

\begin{itemize}
\item $P(y_{2}=16)=0.534135$
\item $P(y_{2}=17)=0.465865$\\
We cannot accept either. (tolerance 60\%)
\end{itemize}
\end{enumerate}
\end{enumerate}
The final answer is therefore $y_{1}=1$ and $y_{2}\in [16,17]$ with
$g(x_{1},x_{2},y_{1},y_{2})=y_{1}x_{1}^{2}+y_{2}x_{2}$.
\end{document}